\documentstyle[12pt]{article}
\def\be{\begin{equation}}
\def\ee{\end{equation}}
\def\ba{\begin{eqnarray}}
\def\ea{\end{eqnarray}}
\def\ga{\mathrel{\raise.3ex\hbox{$>$\kern-.75em\lower1ex\hbox{$\sim$}}}}
\def\la{\mathrel{\raise.3ex\hbox{$<$\kern-.75em\lower1ex\hbox{$\sim$}}}}

\begin{document}

\begin{titlepage}
\pagestyle{empty}
\baselineskip=21pt
\rightline{UMN--TH--1756/99}
\rightline{TPI--MINN--99/18}
\rightline{hep-ph/9903524}
\rightline{March 1999}
\vskip.25in
\begin{center}

{\large{\bf On the Realization of Assisted Inflation}}
\end{center}
\begin{center}
\vskip 0.5in

{Panagiota Kanti and Keith A. Olive}

{\it
{Theoretical Physics Institute, School of Physics and Astronomy, \\
University of Minnesota, Minneapolis, MN 55455, USA}}
\vskip 0.5in
{\bf Abstract}
\end{center}
\baselineskip=18pt \noindent
We consider conditions necessary for a successful implementation of 
so-called assisted inflation.  We generalize the applicability of
assisted inflation beyond exponential potentials as originally proposed
to include standard chaotic ($\lambda \phi^4$ or $m^2 \phi^2$) models as
well. We also demonstrate that in a purely 4-dimensional theory,
unless the assisted sector is in fact decoupled, the additional fields
of the assisted sector actually impede inflation.
As a specific example of an assisted sector, we consider a 
5-dimensional KK model for which the extra dimension may be
somewhat or much larger than the inverse Planck scale.  In this case,
the assisted sector (coming from a KK compactification) eliminates the
need for a fine-tuned quartic coupling to drive chaotic inflation.  This
is a general result of models with one or more ``large" extra
dimensions.   

\vspace*{20mm}
\begin{flushleft}
\begin{tabular}{l} \\ \hline
{\small Emails: yiota@physics.umn.edu, olive@mnhep.hep.umn.edu}
\end{tabular}
\end{flushleft}

\end{titlepage}
\baselineskip=18pt

\section{Introduction}

One of the long standing problems in inflationary model building is the
apparent necessity of some fine-tuned couplings or masses (see
\cite{reviews} for reviews).  Barring an alternative to standard
inflation, either a model predicting the presence of small couplings,
or a more innovative model which does not require them must be
constructed. Developments such as the pre-big bang model \cite{pbb} go a
long way towards this goal, but issues such as the graceful exit still
require resolution \cite{ge}.

The simplest inflationary scenarios are by far the chaotic inflation
models \cite{chaotic} involving a single scalar field. For example, a
potential of the form $V(\phi) = \lambda \phi^4$ will produce sufficient
inflation if the initial background field value is $\phi > $ few $ M_P$.
However, in order to obtain the correct magnitude for density
fluctuations, one must require that the vacuum energy density during the
last $\sim$50 e-foldings of inflation is of order $V \sim (10^{16}~{\rm
GeV})^4$ or $\lambda \sim 10^{-12}$. Similarly chaotic models based on
potentials of the form $V(\phi) = m^2 \phi^2$, require $m \sim 10^{-5}
M_P$ in order to satisfy the COBE constraint.

It is well known that power-law expansion \cite{pl} rather than
exponential expansion may be sufficient to resolve the standard
cosmological problems associated with inflation and that such solutions
can be generated by exponential potentials \cite{pl,exp}.
For example, a potential of the form $V(\phi) = e^{-\lambda \phi}$, leads
to power law expansion with the cosmological scale factor growing as 
$R(t) \sim t^{p}$ with $p = 2/\lambda^2$.  Furthermore, density
fluctuations are no longer scale invariant but scale as 
$|{\delta \rho \over \rho}(k)|^2 \sim k^{n-1}$ with $n= 1 - {2 \over
p-1}$. To obtain, $n \simeq 1$, one requires $p$ to be large.

Recently, it was noticed \cite{liddle} that a system of several scalar
fields each with a potential 
\be
V_i = V_0 e^{-\sqrt{2 \over p_i} \phi_i}
\label{dec}
\ee
could drive power law inflation with a net power $\tilde p = \sum p_i$
sufficient to solve the cosmological problems, even if each of the fields
$\phi_i$ alone are not capable of doing so. Furthermore, the spectral
index of density fluctuations is also brought closer to the scale invariant
spectrum if $p$ is replaced by $\tilde p$.  The dynamics of this type of
``assisted" model was discussed in \cite{mw}. 

Here, we will show that the assisted paradigm can easily be extended to
other types of inflationary models such as the chaotic models mentioned
above.  We will also show that the ansatz (\ref{dec}) of effectively
decoupled scalar fields is absolutely necessary for assistance to work.
For example, the case of $N$ scalar fields each with a potential
defined by $p_i = p$, would lead to $\tilde p = N p$ for self-coupled
fields, while it would lead to $\tilde p = p/N$ for a system of fields
which were cross-coupled. Such a situation would undermine the benefits of
assisted inflation.

Although the identity of these multiple fields was not specified in
\cite{liddle}, one possible source for the necessary multiplicity is a
theory with an extra compact dimension.  The Kaluza-Klein reduction of a
scalar field in 5 dimensions, will result in a spectrum of states with
masses $\propto n^2/L^2$ where $L$ is the size of the compact extra
dimension. If $L \gg M_P^{-1}$, there may be many nearly massless
``copies" of the original scalar field which may serve to assist
inflation. We find that, although the resulting system of scalar fields
produced from the KK reduction may be heavily cross-coupled, it can
eliminate the usual fine-tuning associated with chaotic inflation driven
by a quartic coupling and achieves the goals of assisted inflation.  

\section{Assisted inflation and decoupled fields}

Assisted inflation as described in \cite{liddle,mw} relies on the premise
that there exist a set of $N$ scalar fields each with potential given by
(\ref{dec}).  The Lagrangian for the system is given by
\be
-{\cal L}= \sum_{i=1}^N \Biggl\{ 
\frac 12\,(\partial \phi_i)^2 + V_i \Biggr\}\,\,.
\ee
Each field $\phi_i$ satisfies its equation of motion
\be
\ddot \phi_i + 3 H \dot \phi_i = -{dV_i \over d \phi_i}
\ee
where the Hubble parameter is given by 
\be
3H^2 =  \sum_{i=1}^N \Biggl\{ \frac 12\,\dot \phi_i^2 + V_i \Biggr\}\,\,.
\ee
(We are working in units such that $8 \pi /M_P^2 = 1$.)
In \cite{liddle,mw} it was shown that this system has a late-time
attractor solution described by a single rescaled scalar field 
${\tilde \phi}^2 = (\tilde p / p_1) \phi_1^2$ with potential 
${\tilde V} = (\tilde p / p_1) V_1$.  The resulting power-law expansion
of the Universe is simply $R(t) \sim t^{\tilde p}$ (provided that each of
the $p_i > 1/3$).

As we will now demonstrate, the basic idea behind assisted inflation can
be applied more generally than the case of exponential potentials.  We can
consider a general field theory of multiple, self-interacting scalar
fields of the form
\be
-{\cal L}= \sum_{i=1}^N \Biggl\{ 
\frac 12\,(\partial \phi_i)^2+  \frac{m^2}{2}\,\phi_i^2 \Biggr\} + 
\sum_{i=1}^N \Biggl\{\frac{\lambda_3}{3!}\,\phi_i^3 +
\frac{\lambda_4}{4!}\,\phi_i^4 \Biggl\}\,\,.
\ee
The equation of motion for each field $\phi_i$ derived from the variation
of the above Lagrangian has the form
\be
\nabla^2 \phi_i = m^2\,\phi_i + \frac{\lambda_3}{2}\,\phi_i^2 + 
\frac{\lambda_4}{6}\,\phi_i^3 \,\,.
\ee
{}From the above equation, it is obvious that the system consists
of $N$ completely decoupled scalar fields or equivalently of $N$
copies of the same field. As a result, the Lagrangian can be written as
\ba
-{\cal L} &=& N\,\Biggl\{\frac 12\,(\partial \phi_1)^2 +
\frac{m^2}{2}\,\phi_1^2 + \frac{\lambda_3}{3!}\,\phi_1^3 +
\frac{\lambda_4}{4!}\,\phi_1^4 \Biggr\}
\nonumber \\[2mm]
&=& \frac 12\,(\partial \tilde{\phi})^2 +
\frac{m^2}{2}\,\tilde{\phi}^2 +
\frac{\tilde{\lambda}_3}{3!}\,\tilde{\phi}^3 +
\frac{\tilde{\lambda}_4}{4!}\,\tilde{\phi}^4 \,\,,
\ea
where
\be
\tilde{\phi}=\sqrt{N}\,\phi_1 \quad , \quad
\tilde{\lambda}_3= \frac{\lambda_3}{\sqrt{N}} \quad, \quad
\tilde{\lambda}_4= \frac{\lambda_4}{N} \,\,.
\ee
Notice that the above field redefinition (made to rewrite the Lagrangian
in terms of a field with a canonical kinetic term) results in a scalar
field with an unchanged mass. The resulting theory describes a single
scalar field with the same type of self-inter\-actions compared to the
fields in the original theory. However, these self-interactions are
considerably weaker since both of the coupling constants now scale
with the number of scalar fields
$N$. As a result, as the number of scalar fields that we include in the
theory becomes larger, the effective coupling constants naturally become
smaller and the corresponding fine-tuning becomes milder.
Thus the same basic idea expounded in \cite{liddle,mw} carries over very
simply to chaotic inflation based on a quartic potential. While $\tilde
\lambda_4$ must still be of order $10^{-12}$, the fundamental coupling in
the theory $\lambda_4$ can now be much larger if $N$ is large. Note,
however, that the additional scalar fields do not affect the quadratic
version of chaotic inflation whatsoever.

\section{General theories with cross couplings}

The success of the assisted paradigm demonstrated in the section above,
is directly related to the absence of cross-coupling terms between
different scalar fields. As soon as the multiple self-interacting scalar
fields are substituted with cross-coupled fields, the assistance method
ceases to work. To see that this is the case, it is reasonable to 
consider general field theories of multiple scalar fields of the form
\be
-{\cal L}= \sum_{i=1}^N \Biggl\{ 
\frac 12\,(\partial \phi_i)^2+  \frac{m^2}{2}\,\phi_i^2 \Biggr\} + V_I\,\,,
\ee
where the potential may contain not only self-interaction terms, like the
theory in section 2, but also cross-coupling terms between different fields.
Specifically, we study the following three cases:\\[6mm]
{\bf (A)} We start by considering the following simple theory of coupled
scalar fields with cubic and quartic interaction terms:
\be
V_I =  
\frac{\lambda_3}{3!}\,\Bigl(\sum_{i=1}^N \phi_i\Bigr)^3 +
\frac{\lambda_4}{4!}\,\Bigl(\sum_{i=1}^N \phi_i\Bigr)^4\,\,.
\ee
In this form, the invariance of the theory under the change
$\phi_i \leftrightarrow \phi_j$ is obvious which leads to identical
equations of motion for each of the different scalar fields 
\be
\nabla^2 \phi_i = m^2\,\phi_i + 
\frac{\lambda_3}{2}\,\Bigl(\sum_{k=1}^N \phi_k\Bigr)^2 +
\frac{\lambda_4}{6}\,\Bigl(\sum_{k=1}^N \phi_k\Bigr)^3 \,\,.
\ee
By subtracting the equations of motion of two arbitrary fields $\phi_i$
and $\phi_j$, we can easily see that the solution $\phi_i=\phi_j$ is the
unique late-time attractor of the system. As a result, the Lagrangian
can be written as
\ba
-{\cal L} &=& N\,\Biggl[\frac 12\,(\partial \phi_1)^2 + 
\frac{m^2}{2}\,\phi_1^2\Biggr] + 
\frac{\lambda_3}{3!}\,(N\,\phi_1)^3 + \frac{\lambda_4}{4!}\,
(N\,\phi_1)^4
\nonumber \\[2mm]
&=& \frac 12\,(\partial \tilde{\phi})^2 + \frac{m^2}{2}\,\tilde{\phi}^2+
\frac{\tilde{\lambda}_3}{3!}\,\tilde{\phi}^3 +
\frac{\tilde{\lambda}_4}{4!}\,\tilde{\phi}^4 \,\,,
\ea
where
\be
\tilde{\phi}=\sqrt{N}\,\phi_1 \quad , \quad
\tilde{\lambda}_3= \lambda_3\,N^{3/2} \quad, \quad
\tilde{\lambda}_4= \lambda_4 N^2 \,\,.
\ee
We notice that, when we allow cross-coupling terms between different
fields to be present in the theory, we obtain a result for the effective
potential which is radically different from the one we found in the
case of self-interacting fields in section 2. The presence of these
cross-coupling
terms drives the effective potential, or the coupling constants, in
the opposite direction from that desired: the renormalized, single
scalar field $\tilde{\phi}$ is more strongly coupled than the original
scalar fields $\phi_i$ with the coupling parameters, $\tilde{\lambda}_3$
and $\tilde{\lambda}_4$, increasing with the number of scalar fields that
we include in the theory. As a result, the necessary fine-tuning of the
coupling constants becomes now much more severe.\\[5mm]
{\bf (B)} A slightly different version of the above coupled scalar field
theory can be formulated in the following way. Consider,
\ba
-{\cal L} &=& \sum_{i=1}^N \Biggl\{\frac 12\,(\partial \phi_i)^2+ 
\frac{m^2}{2}\,\phi_i^2 \Biggr\} + 
\sum_{i=1}^N \Biggl\{\frac{\lambda_3}{3!}\,\phi_i^3 +
\frac{\lambda_4}{4!}\,\phi_i^4 \Biggl\} \nonumber\\[2mm]
&+& \sum_{i,j,k=1}^N \frac{\lambda_3c_3}{3!}\,\phi_i \phi_j \phi_k +
\sum_{i,j,k,l=1}^N \frac{\lambda_4 c_4}{4!}\,\phi_i \phi_j \phi_k \phi_l\,\,.
\ea
In the last two terms, the indices ($i, j, k$) and ($i, j, k, l$) are not
allowed to all take on the same value and, as a result, these terms
describe only cross-couplings between different fields. The above
formulation, i.e. the introduction of the parameters $c_3$ and $c_4$ in
the cubic and quartic interaction terms, respectively, allows us to turn
off the cross-couplings between the scalar fields while keeping the
self-interactions in the theory.

We can rewrite the above Lagrangian in the following way
\ba
-{\cal L} &=& \sum_{i=1}^N \Biggl\{\frac 12\,(\partial \phi_i)^2+ 
\frac{m^2}{2}\,\phi_i^2 \Biggr\} + 
\sum_{i=1}^N \Biggl\{\frac{\lambda_3}{3!}\,\phi_i^3 +
\frac{\lambda_4}{4!}\,\phi_i^4 \Biggl\} \nonumber \\[3mm] &+&
\frac{\lambda_3 c_3}{3!}\,\Biggl[\Bigl(\sum_{i=1}^N \phi_i\Bigr)^3-
\sum_{i=1}^N \phi_i^3 \Biggr] 
+ \frac{\lambda_4 c_4}{4!}\,\Biggl[\Bigl(\sum_{i=1}^N \phi_i\Bigr)^4-
\sum_{i=1}^N \phi_i^4 \Biggr] \nonumber\\[4mm]
&=& \sum_{i=1}^N \Biggl\{\frac 12\,(\partial \phi_i)^2+
\frac{m^2}{2}\,\phi_i^2 \Biggr\} + 
\frac{\lambda_3}{3!}\,\Biggl[ (1-c_3)\,\sum_{i=1}^N \phi_i^3 +
c_3\,\Bigl(\sum_{i=1}^N \phi_i\Bigr)^3\,\Biggr]\nonumber\\[3mm]&+&
\frac{\lambda_4}{4!}\,\Biggl[ (1-c_4)\,\sum_{i=1}^N \phi_i^4 +
c_4\,\Bigl(\sum_{i=1}^N \phi_i\Bigr)^4\,\Biggr]
\ea
and, then, the equation of motion for each field $\phi_j$ has the form
\be
\nabla^2 \phi_j = m^2\,\phi_j +
\frac{\lambda_3}{2}\,\Bigl[(1-c_3)\,\phi_j^2+
c_3\,\Bigl(\sum_{i=1}^N \phi_i\Bigr)^2\,\Bigr]+
\frac{\lambda_4}{6}\,\Bigl[ (1-c_4)\,\phi_j^3 +
c_4\,\Bigl(\sum_{i=1}^N \phi_i\Bigr)^3\,\Bigr]\,\,.
\ee
If we subtract the equations of motion of the fields $\phi_j$ and $\phi_k$,
we can easily conclude that, once again, the unique late-time attractor of
the theory has all of the fields equal. By making use of this result, the
Lagrangian may be written as
\ba
-{\cal L} &=& N\,\Biggl[\frac 12\,(\partial \phi_1)^2 +
\frac{m^2}{2}\,\phi_1^2\Biggr] + 
\frac{\lambda_3}{3!}\,\Biggl[(1-c_3)\,N\phi_1^3+ c_3\,(N\phi_1)^3\Biggl]
\nonumber\\[2mm]
&+& \frac{\lambda_4}{4!}\,\Biggl[ (1-c_4)\,N\phi_1^4 + c_4\,(N\phi_1)^4\Biggr]
\nonumber \\[3mm]
&=& \frac 12\,(\partial \tilde{\phi})^2 + \frac{m^2}{2}\,\tilde{\phi}^2 +
\frac{\tilde{\lambda}_3}{3!}\,\tilde{\phi}^3\,\Bigl[1+c_3\,(N^2-1)\Bigr]+
\frac{\tilde{\lambda}_4}{4!}\,\tilde{\phi}^4\,\Bigl[1+c_4\,(N^3-1)\Bigr], 
\ea
where
\be
\tilde{\phi}=\sqrt{N}\,\phi_1 \quad , \quad
\tilde{\lambda}_3= \frac{\lambda_3}{\sqrt{N}}\quad, \quad
\tilde{\lambda}_4= \frac{\lambda_4}{N}\,\,.
\label{good2}
\ee
\paragraph{}
If we choose $c_3=c_4=0$, then, we recover the theory of self-interacting
scalar fields that was discussed in section 2 and for which the assistance
effect worked perfectly leading to an extremely weakly coupled scalar field
theory. If, on the other hand, we choose $c_3=c_4=1$, then, we go back to
the case {\bf (A)} studied above, where the potential increases rapidly with
the number $N$ of scalar fields. A third  possibility
arises when the parameters $c_i$ adopt some intermediate values. For
example, if, for large $N$, $c_3 \sim 1/N^2$ and $c_4 \sim 1/N^3$, the
coefficients of the renormalized cubic and quartic terms that appear inside
the brackets are of ${\cal O}(1)$
and the desired behavior (\ref{good2}) of the coupling parameters
$\tilde{\lambda}_i$ is ensured. One could argue that the result of this
analysis is to transfer the fine-tuning from the coupling constants to
the parameters $c_i$. Indeed, it shows the degree to which the
cross-couplings must be fine-tuned for assistance to work.

\noindent {\bf (C)} Finally, we consider a theory of $N$ scalar fields
coupled to each other through an exponential potential
\be
-{\cal L}=\sum_{i=1}^N\,\frac 12\,(\partial\phi_i)^2 +
V_0 \,\prod_{i=1}^N \,\exp\Biggl(-\sqrt{\frac{2}{p}}\,
{\phi_i}\Biggr)\,\,.
\label{trans}
\ee
This is similar to the potential considered by Liddle et
al. \cite{liddle}  with the sum of exponentials replaced by a product. In
the case of the summation, the $N$ scalar fields do not interact with
each other and the unique late-time attractor has all fields equal.
As discussed in the introduction, this solution leads to a power law
expansion which can solve the standard inflationary problems with a
relatively flat spectrum of density fluctuations.

In our case, however, the scalar fields are coupled to
each other.  The equation of motion for the field $\phi_i$
takes the form
\be 
\nabla^2 \phi_i = -\sqrt{\frac{2}{p}}\,{V_0}\,
\exp\Biggl(-\sqrt{\frac{2}{p}}\,\sum_{k=1}^N
{\phi_k}\Biggr)\,\,.
\ee
The right-hand-side of the above equation is the same for every field
$\phi_i$. As a result, the unique late-time attractor, which has all the
fields equal, is still valid even in this case where the fields are coupled.
Now, the Lagrangian can be written as
\be
-{\cal L}=\frac N2\,(\partial\phi_1)^2 +
V_0 \,\exp\Biggl(-\sqrt{\frac{12}{p}}\,
{N\,\phi_1}\Biggr)=\frac 12\,(\partial\tilde{\phi})^2 +
V_0 \,\exp\Biggl(-\sqrt{\frac{2}{\tilde{p}}}\,
{\tilde{\phi}}\Biggr)
\ee
where now
\be
\tilde{\phi}=\sqrt{N}\,\phi_1 \quad , \quad \tilde{p}=\frac{p}{N}
\ee
As a result, if $p$, in the original theory, was not large enough to
support inflation, the situation is worsened  since $p$ is divided by
the number of scalar fields that are present in the theory. 

In each of the cases studied above, it is evident that the presence
of interaction terms between the scalar fields of the theory undermines
the benefits of assistance and impedes the successful
implementation of inflation.  
While the cases we studied are certainly simplified, we expect the
general result to hold, namely, in a theory with multiple scalar fields,
assistance requires the absence (or near absence) of cross-couplings
between the scalar fields.
\section{Field theories with multiple scalar fields}

Given the potential utility of having several or many fields which are in
some sense copies of each other, we now look at a possible source for
these fields in theories with extra spatial dimensions. It is well known
that the Kaluza-Klein reduction of a theory leads to the existence of
many new fields which appear as zero-modes in the final 4-dimensional
theory. For example, consider a simple 5-dimensional gravitational action
of the form
\be
 S_G= -\int d^{5}x \sqrt{G_5} \Bigl\{{M_5^3 \over 16 \pi}
\,R_5  \Bigr\} \label{act5} \ee 
where $ M_5$ is the five-dimensional Planck mass. Upon compactification
along dimension of circumference $2 L$, we obtain
\be
S_G  =-\frac12\int d^{4}x \sqrt{G_4} e^\gamma \Bigl\{R_4 +
e^{2\gamma} {1 \over 4}\,F_{KK}^2 \Bigr\}
\label{act4}
\ee
where $M_P^2 = 2 L M_5^3$, $\gamma$ is the scalar associated with the
5-5 component of the metric ($e^{2\gamma} = g_{55}$), and $F_{KK}$
is the field strength of the Kaluza-Klein gauge field associated with
$g_{\mu5}$. This action can be brought into the Einstein frame by the
conformal transformation ${G_4}_{\mu\nu} = e^{-\gamma} g_{\mu\nu}$
to give
\be
S_G  =-\frac12\int d^{4}x \sqrt{g} \Bigl\{R + {3\over2}\,(\partial \gamma)^2
+ e^{3\gamma} {1 \over 4}\,F_{KK}^2 \Bigr\}\,\,.
\label{act4e}
\ee

Let us further suppose that the original 5-dimensional theory contains an
additional massless scalar field $\hat \phi$ with action
\be
S_\phi= -\int d^{5}x \sqrt{G_5} \Bigl\{{G_5}^{AB} \partial_A \hat \phi
\,\partial_B \hat \phi \Bigr\}
\label{actphi5}
\ee
where the indices $A, B =\{t,x_1,x_2,x_3,z\}$. We can Fourier expand
$\hat \phi$ along $z$ as
\be
\hat \phi(x,z) = \hat \phi_0(x) + \hat \phi_z(x,z)= \hat \phi_0(x)
+ \sum_{n = 1}^{\infty} 
\,\Bigl[ \hat \phi_n(x)\,e^{i\frac{n\pi}{L}z} + 
\hat \phi^{*}_n(x)\,e^{-i \frac{n\pi}{L}z} \Bigr]\,\,,
\label{expansion}
\ee
where $\hat \phi_0$ is the 5-dimensional field that depends only on 
non-compact coordinates. 

Upon reducing to 4 dimensions, and performing the same conformal
transformation, the action (\ref{actphi5}) becomes
\be
S_\phi= -\int d^{4}x \sqrt{g}\,\Bigl\{\sum_{n = 0}^{\infty} \Bigl(
|(\partial_\mu + i{n\pi \over L} A_\mu) \phi_n|^2 + {n^2 \pi^2
\over L^2}e^{-3\gamma} |\phi_n|^2 \Bigr)
\Bigr\}
\label{actphi4}
\ee
where we have defined the 4-dimensional scalar field $\phi = \sqrt{2L}
\,\hat \phi$.  In what follows,  we will assume that the dilaton-like
field $\gamma$ is fixed \cite{ac}, and ignore the role of the KK gauge
field $A_\mu$. Although we have written the action in terms of an infinite
sum, the momentum along $z$, $p_z$, should be limited by $M_5$. In that
case, we should only consider fields up to $n = N \la L M_5/\pi$. For
$(\pi L^{-1}
\ll M_5)$, there may be many fields which can  in principle assist
inflation.  Such theories are, to say the least, of wide interest at the
moment \cite{large} (see also \cite{nem} and references therein).

Let us now consider the following
5-dimensional scalar field, self-interacting through a quartic
potential, as a concrete example: 
\be
-{\cal L}_{5D}= \frac 12\,\partial_A \hat \phi\,\partial^A \hat \phi +
\frac{\hat \lambda}{4!M_5}\,\hat \phi^4\,\,.
\ee
The kinetic term for the 5-dimensional field $\hat \phi$ can be expanded
as in (\ref{actphi4}). Similarly, the substitution of the expansion
(\ref{expansion}) in the potential gives rise to numerous interaction terms
between the Kaluza-Klein scalar fields. Then, the 4-dimensional Lagrangian
can be written as
\ba
-{\cal L}_{4D} &=& 
\frac 12\,\partial_\mu \phi_0\,\partial^\mu \phi_0 + 
\sum^{\infty}_{n=1} \Bigl(\partial_\mu \phi_n \partial^\mu \phi_n^* +
\frac{n^2 \pi^2}{L^2}\,\phi_n \phi_n^* \Bigr) \nonumber \\[1mm]
&+& \frac{\lambda}{4!}\,\Biggl[\phi^4_0 + 12 \phi^2_0 
\sum_{n=1}^{\infty} \phi_n \phi_n^*
+ 12 \phi_0 \sum_{n,k=1}^{\infty} 
\Bigl(\phi_n \phi_k \phi^*_{n+k}
+ \phi^*_n \phi^*_k \phi_{n+k} \Bigr)\nonumber \\[1mm]
&+& 
\sum_{n,k,l =1}^{\infty} 
\Bigl( 4 \phi_n \phi_k \phi_l \phi^*_{n+k+l}
+ 4 \phi^*_n \phi^*_k \phi^*_l \phi_{n+k+l} +
6 \phi_n \phi_k \phi^*_l \phi^*_{n+k-l (\ge 1)} \Bigr)\Biggr]\,\,,
\label{lagrangian}
\ea
in terms of the 4-dimensional field $\phi$ and the 4-dimensional
coupling $\lambda=\hat{\lambda}/(2 L M_5)$. 
Note that in the last term in the last equation we include only
the terms for which $n+k-l \ge 1$. It is also important to note that the
4-dimensional (dimensionless) coupling is now reduced relative to the
original 5-dimensional coupling $\hat \lambda$ by an amount $2LM_5
\simeq N$. 

It will be useful to begin the analysis of this system by first 
simplifying to a restricted set of fields.  Thus, in the lowest order
approximation, we may assume that, apart from the field $\phi_0$, only
$\phi_1$ and $\phi_1^*$ are present in the theory and  set all the other
Kaluza-Klein fields equal to zero. By making use of the definitions 
\be 
\phi_1 = \frac{X + i Y}{\sqrt{2}} \qquad , \qquad 
m^2=\frac{\pi^2}{L^2}\,\,,
\ee
the effective Lagrangian takes the form
\ba
-{\cal L}_{eff} &=& \frac12\,(\partial \phi_0)^2
+ \frac12\,(\partial X)^2 + \frac12\,(\partial Y)^2 
+ \frac{m^2}{2}\,(X^2 + Y^2) \nonumber \\[2mm]
&+& \frac{\lambda}{4!}\,\phi_0^4 + \frac{\lambda}{4} 
\,\phi_0^2\,(X^2 + Y^2) + \frac{\lambda}{16}\,(X^2 + Y^2)^2\,\,.
\ea
The variation of this Lagrangian with respect to $\phi_0$, $X$
and $Y$ leads to the following equations of motion
\ba
\nabla^2 \phi_0 &=& \frac{\lambda}{6}\,\phi_0^3 + \frac{\lambda}{2}
\,\phi_0\,(X^2 + Y^2) \nonumber \\[2mm]
\nabla^2 X &=& m^2 X + \frac{\lambda}{2}\,\phi_0^2\,X 
+ \frac{\lambda}{4}\,X (X^2+Y^2) \nonumber \\[2mm]
\nabla^2 Y &=& m^2 Y + \frac{\lambda}{2}\,\phi_0^2\,Y 
+ \frac{\lambda}{4}\,Y (X^2+Y^2)\,\,. 
\ea
Obviously, the latter two equations are the same, and so we can set
\be
Y = \kappa X \label{assum1}\,\,.
\ee
Making this substitution for $Y$, the first
two equations of motion reduce to
\ba
\nabla^2 \phi_0 &=& \frac{\lambda}{6}\,\phi_0^3 + \frac{\lambda}{2}
\,\phi_0\,X^2 (1 + \kappa^2) \nonumber \\[2mm]
\nabla^2 X &=& m^2 X + \frac{\lambda}{2}\,\phi_0^2\,X
+ \frac{\lambda}{4}\,X^3 (1+\kappa^2)\,\,. 
\ea
As long as the mass term is negligible
compared to the cubic term, i.e. when $m^2 << \lambda\,\phi_0^2/2$, 
\be
\phi_0 = q X \label{assum2}
\ee
is also a solution, provided that
\be
q^2 = \frac34\,(1 + \kappa^2)\,\,.
\label{cona1}
\ee
In this case, the kinetic part of the Lagrangian can be written as
\be
-{\cal L}_{kin} = \frac12\,(\partial \phi_0)^2 + 
\frac12\,(\partial X)^2 + \frac12\,(\partial Y)^2
= \frac12\,(\partial \phi_0)^2 (1 + \frac{1+\kappa^2}{q^2})
\ee
and by using the constraint (\ref{cona1}), we obtain
\be
-{\cal L}_{kin} = \frac12\,\frac73\,(\partial \phi_0)^2=
\frac12\,(\partial \tilde{\phi})^2\,\,,
\ee
where we have implemented the field redefinition
\be
\tilde{\phi} = \sqrt{\frac73}\,\phi_0
\ee
in order to map the system of the three real, scalar fields to a
theory of a single scalar field. Next, we look at the quartic potential
which now takes the form
\be
V_{eff} = \frac{\lambda}{4!} \phi_0^4 +
\frac{\lambda}{4}\,\phi_0^2\,(X^2 + Y^2) 
+ \frac{\lambda}{16}\,(X^2 + Y^2)^2=
\frac{35}{3}\,\frac{\lambda}{4!}\,\phi_0^4 = 
\frac{15}{7}\,\tilde{V}\,\,,
\label{veffn=1}
\ee
where $\tilde{V}$ is the quartic potential of the renormalized scalar
field $\tilde{\phi}$. As a result, the presence of the two Kaluza-Klein
fields $X$ and $Y$ have raised the value of the effective potential by a
factor of $15/7$ and so, the fine tuning of the coupling constant $\lambda$
has worsened.
Note that this result does not depend on the substitution
(\ref{assum1}) and holds even if we set
$Y=0$.

Before we tackle the more general case, it will still be useful to
consider the next-to-lowest order where we allow 
$\phi_2$ and $\phi_2^*$, apart from $\phi_0$, $\phi_1$ and $\phi_1^*$, 
to be present in the theory. In a similar way, we set
\begin{equation}
\phi_1=\frac{X_1+iY_1}{\sqrt{2}} \qquad, \qquad 
\phi_2=\frac{X_2+iY_2}{\sqrt{2}} \,\,,
\end{equation}
and define
\begin{equation}
m_1^2=\frac{\pi^2}{L^2} \qquad, \qquad m_2^2=\frac{4 \pi^2}{L^2}\,\,.
\end{equation}
\paragraph{}
Once again the system can be simplified. In this case,
the choice $X_2=0$ corresponds to a special solution of the five field
system  as long as $X_1^2=Y_1^2$. Then, the remaining equations of motion
are given by
\begin{eqnarray}
\nabla^2 \phi_0 &=& \frac{\lambda}{6}\,\phi_0^3 + \frac{\lambda}{2}
\,\phi_0\, (2 X_1^2 + Y_2^2) +
\frac{\lambda}{\sqrt{2}}\,X_1 Y_1 Y_2
\label{a} \\[2mm]
\nabla^2 Y_1 &=& m_1^2\,Y_1 + \frac{\lambda}{2}\,\phi_0^2\,Y_1
+\frac{\lambda}{\sqrt{2}}\,\phi_0\,(X_1 Y_2)
+ \frac{\lambda}{2}\,Y_1\,(X_1^2 + Y_2^2)
\label{b} \\[2mm]
\nabla^2 Y_2 &=& m_2^2\,Y_2 + \frac{\lambda}{2}\,\phi_0^2\,Y_2
+\frac{\lambda}{\sqrt{2}}\,\phi_0\,(X_1 Y_1)
+ \frac{\lambda}{4}\,Y_2\,(4 X_1^2 + Y_2^2)
\label{c}
\end{eqnarray}

If we further assume, as before, that the masses $m_1^2 $ and
$m_2^2$ are small compared to $\lambda \phi^2/2$ and neglect them, the
ansatz
\begin{equation}
\phi_0=q X_1 \qquad , \qquad Y_2=p\,Y_1 
\label{spec}
\end{equation}
is indeed a solution of the system (\ref{a})-(\ref{c}). Rearranging
equations (\ref{a})-(\ref{b}) and (\ref{b})-(\ref{c}), we obtain the
constraints
\ba
&~& 3 \sqrt{2} p\,(1-q^2) - 2 q^3 + 3q = 0\,\,, \label{conb1}\\[2mm]
&~& 2 \sqrt{2} q\,(1-p^2) - p^3 + 2p = 0\,\,,
\label{conb2}
\ea
respectively. The above system of algebraic equations can be solved
numerically leading to the following pairs of values for the
proportionality coefficients $q$ and $p$
\begin{eqnarray}
&(A)&: \,\,q \rightarrow 2.22 \qquad \hspace*{0.22cm}, 
\qquad p \rightarrow -0.911 \label{A} \\[2mm]
&(B)&: \,\,q \rightarrow 1.07 \qquad \hspace*{0.22cm},
\qquad p \rightarrow  1.13 \label{B} \\[2mm]
&(C)&: \,\,q \rightarrow 0.956 \qquad , \qquad p \rightarrow -3.07 
\label{C} \\[2mm]
&(D)&: \,\,q \rightarrow 0.722 \qquad , \qquad p \rightarrow -0.695
\label{D}
\end{eqnarray}
The above set of solutions are supplemented by another set where the
signs of $q$ and $p$ are opposite. But, as we will see,
both the potential and the kinetic terms are invariant under the 
simultaneous change $q \rightarrow -q$ and $p \rightarrow -p$
and so, we may ignore the second set of solutions.

By setting $X_2=0$ and using the proportionality relations that hold between
the remaining four scalar fields, the kinetic Lagrangian takes the form
\begin{eqnarray}
-{\cal L}_{kin} &=&
\frac 12 (\partial \phi_0)^2 + \frac 12 (\partial X_1)^2 +
\frac 12 (\partial Y_1)^2 + \frac 12 (\partial Y_2)^2
\nonumber \\[2mm]
&=& \frac 12 \,(\partial \phi_0)^2\,\Bigl(1+\frac{2+p^2}{q^2}\Bigr)
= \frac 12 \,(\partial \tilde{\phi})^2
\end{eqnarray}
where
\begin{equation}
\tilde{\phi}= \phi_0\,\sqrt{1+\frac{2+p^2}{q^2}}\,\,.
\label{renorm}
\end{equation}
In the same way, the effective potential can be written as
\begin{eqnarray}
V_{eff} &=& \frac{\lambda}{4!}\,\tilde{\phi}^4\,\Biggl\{ 1 + 
\frac{6(2+p^2)}{q^2} + \frac{12 \sqrt{2} p}{q^3} +
\frac{3}{2q^4}\,(4+8p^2+p^4)\Biggr\}\,
\Biggl(1+\frac{2+p^2}{q^2}\Biggr)^{-2}\,\,.
\end{eqnarray}
Substituting the values of the parameters $q$ and $p$ from solutions
(\ref{A})-(\ref{D}), we obtain the final results for the value of the
effective potential
\begin{eqnarray}
&(A)& : \,\,V_{eff}=1.51\,\tilde{V} \label{res1}\\[2mm]
&(B)& : \,\,V_{eff}=3.48\,\tilde{V} \label{veffn=2}\\[2mm] 
&(C)& : \,\,V_{eff}=1.75\,\tilde{V} \label{res3}\\[2mm]
&(D)& : \,\,V_{eff}=1.29\,\tilde{V}\label{res4}
\end{eqnarray}
where the renormalization of the scalar field (\ref{renorm}) has been 
taken into account. According to the above results, there is one solution
that multiplies the one-field-potential $\tilde{V}$ by a nume\-rical
coefficient 3.48, which can be compared to the lowest order solution
where the potential was multiplied by $15/7 \simeq$ 2.14. In this case
the fine-tuning is further aggravated. However, there are three additional
solutions that multiply the potential by coefficients which, although
larger  than unity, are smaller than the first one. Thus, there is the
hope that as we add more and more extra fields these coefficients become
smaller. Recall, that our goal is to ease the original fine-tuning
problem associated with (in this case) a simple $\lambda \phi^4$ chaotic
inflationary model.  Furthermore, for $N
\sim 2LM_5 \gg 1$, the 4-dimensional coupling is $\lambda \sim \hat
\lambda /N$ and will be small provided, $ \hat \lambda \sim 1$ and $N \ga
10^{12}$ or equivalently, $M_5 \la 10^{-6} M_P$. 
Therefore, so long as the potential does not grow as $N$, there will be a
viable solution for the assisted paradigm.

One could argue that the above results for the effective potential hold only
for the special solution (\ref{spec}) supplemented by the relations $X_2=0$
and $X_1^2=Y_1^2$ that we have considered. For this reason, we studied
some additional special solutions of the equations of motion, namely
\ba
&~& \phi_0=q X_1 \quad , \quad X_2=p\,X_1 \quad , \quad Y_1=Y_2=0 \\[2mm]
&~& \phi_0=q Y_1 \quad , \quad X_2=p\,Y_1 \quad , \quad X_1=Y_2=0\,\,.
\ea
Under a numerical renormalization of the proportionality coefficients, i.e.
\be
q \rightarrow \pm\frac{\tilde{q}}{\sqrt{2}} \qquad , \qquad
p \rightarrow \frac{\tilde{p}}{\sqrt{2}}\,\,,
\ee
the above solutions, substituted in the equations of motion, lead to the
same constraints (\ref{conb1})-(\ref{conb2}) for the coefficients
$\tilde{q}$ and $\tilde{p}$ and the same results
(\ref{res1})-(\ref{res4}) for the effective potential. As in the lowest
order, the effective potential seems to depend merely on the number of
the real scalar fields included in the theory and not on their ``flavor",
i.e. if they come from the real or imaginary parts of the complex
Kaluza-Klein fields or from an arbitrary combination of them. 

Indeed, the invariance of the effective potential under the
selection of different special solutions exists, as long as these
solutions are characterized by the same number of real fields and lead to
the same number of constraints on the proportionality coefficients
involved. As we mentioned above, the result is always independent of the
origin of these real fields. We may, then,
conclude, that the real and imaginary parts of $N$ Kaluza-Klein complex
fields, that come from the compactification of the fifth dimension,
constitute equivalent degrees of freedom contributing equally to the
final number of $2N$ degrees of freedom. This result allows us to
substitute the $N$ complex Kaluza-Klein fields with $2N$ real fields. 
Then, the Lagrangian (\ref{lagrangian}) reduces to
\ba
-{\cal L}_{4D} &=& 
\frac 12\,(\partial \phi_0)^2+ \sum^{2N}_{n=1} 
\Biggl\{\frac 12\,(\partial\phi_n)^2 +
\frac{n^2 \pi^2}{2L^2} \phi_n^2\Biggr\}
+ \frac{\lambda}{4!}\,\phi^4_0 
+ \frac{\lambda}{4}\,\phi^2_0 \sum_{n=1}^{2N} \phi_n^2 \nonumber \\[3mm]
&+& \frac{\lambda}{2\sqrt{2}}\,\phi_0 \sum_{n,k=1}^{2N} 
\phi_n \phi_k \phi_{n+k}
+ \frac{\lambda}{12}\,\sum_{n,k,l=1}^{2N} 
\phi_n \phi_k \phi_l \Bigl(\phi_{n+k+l}
+ \frac 34\,\phi_{n+k-l} \Bigr)
\label{lagr2}
\ea 
with equations of motion given by
\ba
\nabla^2 \phi_0 &=& \frac{\lambda}{6}\,\phi_0^3 + 
\frac{\lambda}{2}\,\phi_0 \sum_{n=1}^{2N} \phi_n^2
+ \frac{\lambda}{2 \sqrt{2}} \sum_{n,k=1}^{2N} 
\phi_n \phi_k \phi_{n+k} \label{eq0} \\[6mm]
\nabla^2\phi_n &=& m_n^2 \phi_n + \frac{\lambda}{2}\,\phi_0^2 \phi_n
+ \frac{\lambda}{2 \sqrt{2}} \phi_0 \sum_{k=1}^{2N}
\Bigl(2\phi_k \phi_{k+n} + \phi_k \phi_{n-k}\Bigr) \nonumber \\[2mm]
&+& \frac{\lambda}{4} \sum_{k,l=1}^{2N} \Biggl\{ \phi_k \phi_l
\Bigl(\phi_{k+l+n} + \frac 13 \phi_{n-k-l} + \frac 34 \phi_{n+k-l}
+ \frac 14 \phi_{k+l-n}\Bigr) \Biggr\}\,\,.
\label{eqn}
\ea
It is to be understood that the fields denoted by index combinations
such as $n+k+l$ and $n+k-l$ are to be included only if they are $\ge 1$
and $\le 2N$.
 
Although the substitution of complex fields by an equivalent number of
real fields simplifies the theory, it does not allow us to study the
effect from the addition of a large number of extra scalar fields on the
effective potential by using the method described above. The appearance
of new cross-coupling terms as we increase the order of the theory makes
the analytical formulation of the problem extremely tedious and the
determination  of the result for the effective potential impossible even
through numerical methods. 
 
As an alternative approach to the problem, we construct the function
$\phi_{n+1}-\phi_n$ out of the difference of two consecutive Kaluza-Klein
fields and it is easy to show that it satisfies the following equation
\vspace*{1mm}
\ba
\nabla^2(\phi_{n+1}-\phi_n) &=&  
\frac{\lambda}{2}\,\phi_0^2 (\phi_{n+1}-\phi_n)
+ \frac{\lambda}{2 \sqrt{2}}\,\phi_0 \sum_{k=1}^{2N}\Biggl\{
2\phi_k (\phi_{k+n+1}-\phi_{k+n}) \nonumber \\[3mm]
&+ &\phi_k (\phi_{n+1-k}-\phi_{n-k})\Biggr\}
+ \frac{\lambda}{4} \sum_{k,l=1}^{2N} \Biggl\{ \phi_k \phi_l
\Bigl[(\phi_{k+l+n+1}-\phi_{k+l+n}) \nonumber \\[3mm]
&+& \frac 13 (\phi_{n+1-k-l}-\phi_{n-k-l})
+ \frac 34 (\phi_{n+1+k-l}-\phi_{n+k-l})\nonumber \\[3mm]
&+& \frac 14 (\phi_{k+l-n-1}-\phi_{k+l-n}\Bigr] \Biggr\}\,\,.
\ea
The right-hand-side of the above equation, which is proportional to
the first derivative of the effective potential with respect to the field
$\phi_{n+1}-\phi_n$, has a minimum when $\phi_n$ approaches
both $\phi_{n-1}$ and $\phi_{n+1}$ at late times. As a result, one of the
possible late-time attractors of the theory has all of the extra fields
equal. By setting $\phi_1=\phi_2=...=\phi_{2N}$, the calculation of
the effective potential in the presence of $2N$ extra scalar fields
in the theory can be easily conducted.

However, the above argument suffers from two major loopholes: first,
the attractor that has all of the Kaluza-Klein fields equal is only one
of the possible late-time attractors and may be not the one chosen by
the system and, second, the condition
$\phi_{n-1}=\phi_n=\phi_{n+1}$ can not be fulfilled for the ``boundary 
fields" $\phi_1$ and $\phi_{2N}$. In the case $n=1$, the field $\phi_{n-1}$
does not exist by construction and the same holds for $\phi_{n+1}$ when
$n=2N$. Both of the above problems can be eliminated by imposing the
periodic condition $\phi_{2N+i}=\phi_i$ when $2N$ real Kaluza-Klein
fields are present in the theory. Then, the ``boundaries" are removed and
we can define both $\phi_{n-1}$ and $\phi_{n+1}$ for every field $\phi_n$.
Moreover, we may prove that, after the imposition of the periodic condition,
the attractor that has all of the fields equal is the unique late-time
attractor of the system. For this purpose, we are going to make use of
the induction method. We start with the case with 2 real Kaluza-Klein
fields for which the Lagrangian (\ref{lagr2}) becomes
\ba
-{\cal L}_{4D} &=& 
\frac 12\,(\partial \phi_0)^2+ \frac 12\,(\partial\phi_1)^2 +
\frac 12\,(\partial\phi_2)^2 +\frac{\lambda}{4!}\,\phi^4_0 
+ \frac{\lambda}{4}\,\phi^2_0 \,(\phi_1^2+\phi_2^2)\nonumber\\[2mm] 
&+& \frac{\lambda}{2\sqrt{2}}\,\phi_0\,\phi_2\,(3\phi_1^2+\phi_2^2) 
+\frac{7\lambda}{48}\,(\phi_1^4+\phi_2^4+6\phi_1^2\phi_2^2)\,\,.
\ea 
Note that, strictly speaking, the above Lagrangian should follow from
eq.~(\ref{lagrangian}) in the next-to-lowest order considered above
if we put $Y_1=Y_2=0$. However, this is not exactly the case: due to
the boundary condition imposed, there are {\em additional} terms present
in the Lagrangian which modify the coefficients of the cross-coupling terms
while leaving their structure unchanged. The equations of motion of the
fields $\phi_1$ and $\phi_2$, then, have the form
\ba
\nabla^2\phi_1 &=& \frac{\lambda}{2}\,\phi_0^2\phi_1 +
\frac{3\lambda}{\sqrt{2}}\,\phi_0\phi_1\phi_2 +
\frac{7\lambda}{12}\,\phi_1\,(\phi_1^2+3\phi_2^2)\,\,, \\[3mm]
\nabla^2\phi_2 &=& \frac{\lambda}{2}\,\phi_0^2\phi_2 +
\frac{3\lambda}{2\sqrt{2}}\,\phi_0\,(\phi_1^2+\phi_2^2) +
\frac{7\lambda}{12}\,\phi_2\,(\phi_2^2+3\phi_1^2)\,\,.
\ea
Subtracting the above equations, we obtain the result
\ba
\nabla^2(\phi_2-\phi_1) &=& \frac{\lambda}{2}\,\phi_0^2\,(\phi_2-\phi_1) 
+ \frac{3\lambda}{2\sqrt{2}}\,\phi_0\,(\phi_2-\phi_1)^2 +
\frac{7\lambda}{12}\,(\phi_2-\phi_1)^3 \nonumber\\[3mm]
&=& \frac{\lambda}{2}\,\psi\,\Biggl(\phi_0^2 +
\frac{3}{\sqrt{2}}\,\phi_0\psi+\frac{7}{6}\,\psi^2\Biggr)\,\,.
\label{n2}
\ea
The right-hand-side of the above equation is the first derivative of the
effective potential with respect to the field $\psi=\phi_2-\phi_1$. It is
obvious that the choice $\psi=0$ minimizes the potential. Actually, this
is the only minimum of the potential since the expression inside the
brackets does not have any real solutions.

Next, we assume that the only minimum of the effective potential, when
$2N-1$ scalar fields are included in the theory, corresponds to
$\phi_1=\phi_2=...=\phi_{2N-1}$. We will show that if we add one more
field, $\phi_{2N}$, the aforementioned late-time attractor expands in order
to include $\phi_{2N}$, too. So, assuming that we have $2N-1$ equal scalar
fields and the field $\phi_{2N}$, the Lagrangian (\ref{lagr2}) takes the form
\ba
-{\cal L}_{4D} &=& 
\frac 12\,(\partial \phi_0)^2+ \frac{(2N-1)}{2}\,(\partial\phi_1)^2 +
\frac 12\,(\partial\phi_{2N})^2 +\frac{\lambda}{4!}\,\phi^4_0 
+ \frac{\lambda}{4}\,\phi^2_0 \,\Bigl[(2N-1)\phi_1^2+\phi_{2N}^2\Bigr]
\nonumber\\[2mm] 
&+& \frac{\lambda}{2\sqrt{2}}\,\phi_0\,\Biggl\{2(2N-1)(N-1)\phi_1^3+
\phi_{2N}\,\Bigl[3(2N-1)\phi_1^2+\phi_{2N}^2\Bigr]\Biggr\} \nonumber\\[2mm]
&+&\frac{7\lambda}{48}\,\Biggl\{4(2N-1)\,\Bigl[(N-1)\phi_1^2+
\phi_1\phi_{2N}\Bigr]^2+\Bigl[(2N-1)\phi_1^2+\phi_{2N}^2\Bigr]^2\Biggr\}\,\,.
\ea 
Now, the equations of motion of the fields $\phi_1$ and $\phi_{2N}$ take
the form
\ba
\nabla^2\phi_1 &=& \frac{\lambda}{2}\,\phi_0^2\phi_1 +
\frac{3\lambda}{\sqrt{2}}\,\phi_0\,\Bigl[(N-1)\phi_1^2 +
\phi_1\phi_{2N}\Bigr] \nonumber\\[2mm]
&+&\frac{7\lambda}{12}\,\Biggl\{\Bigl[4(N-1)^2+(2N-1)\Bigr]\,\phi_1^3+
6(N-1)\,\phi_1^2\phi_{2N}+3\phi_1\phi_{2N}^2\Biggl\}\,\,, \\[5mm]
\nabla^2\phi_{2N} &=& \frac{\lambda}{2}\,\phi_0^2\phi_{2N} +
\frac{3\lambda}{2\sqrt{2}}\,\phi_0\,\Bigl[(2N-1)\phi_1^2+\phi_{2N}^2\Bigr]
\nonumber\\[2mm] &+& \frac{7\lambda}{12}\,\Bigl[2(2N-1)(N-1)\phi_1^3+
3(2N-1)\phi_1^2\phi_{2N}+\phi_{2N}^3\Bigr]\,\,,
\ea
while the equation of motion of the field $\psi=\phi_{2N}-\phi_1$ is found
to be
\be
\nabla^2(\phi_{2N}-\phi_1) = \frac{\lambda}{2}\,\phi_0^2\,(\phi_{2N}-\phi_1) 
+ \frac{3\lambda}{2\sqrt{2}}\,\phi_0\,(\phi_{2N}-\phi_1)^2 +
\frac{7\lambda}{12}\,(\phi_{2N}-\phi_1)^3\,\,.
\ee
The above equation is identical with eq.~(\ref{n2}) and, as a result, the
effective potential has a unique minimum at $\psi=0$. According to the
above result, the only late-time attractor for the system of $2N$ Kaluza-Klein
scalar fields corresponds to $\phi_1=\phi_2=...=\phi_{2N-1}=\phi_{2N}$.
\paragraph{}
Now, we proceed to calculate the kinetic term and the effective potential
of the system. By using the late-time attractor of equal fields in the
case of $2N$ scalar fields, the kinetic part of the Lagrangian (\ref{lagr2})
takes the form
\be
-{\cal L}_{4D} = \frac 12\,(\partial \phi_0)^2+ 
\frac{(2N)}{2}\,(\partial\phi_1)^2 = \frac 12\,(\partial \phi_0)^2\,
\Bigl(1+\frac{2N}{q^2}\Bigr)=\frac 12\,(\partial \tilde{\phi})^2\,\,,
\ee 
where we have assumed the proportionality relation $\phi_0=q\,\phi_1$
and renormalized the scalar field $\phi_0$. Then, the effective potential
reduces to
\ba
V_{eff} &=& \frac{\lambda}{4!}\,\phi^4_0 + \frac{\lambda}{4}\,(2N)\,
\phi^2_0 \phi_1^2 + \frac{\lambda}{2\sqrt{2}}\,(2N)^2\phi_0\,\phi_1^3+
\frac{7\lambda}{48}\,(2N)^3\phi_1^4 \nonumber\\[3mm]
&=& \tilde{V}\,\Biggl(1+\frac{12N}{q^2}+\frac{24\sqrt{2}N^2}{q^3}
+ \frac{28N^3}{q^4}\Biggr)\,\Biggl(1+\frac{2N}{q^2}\Biggr)^{-2}\,\,,
\label{vfinal}
\ea
where $\tilde{V}$ is the quartic potential of the renormalized field
$\tilde{\phi}$. According to the above result, the effective potential 
depends on two parameters: the number $N$ of Kaluza-Klein fields that
we include in the theory and the proportionality coefficient $q$. This
coefficient, although a number, may itself depend on $N$ changing radically
the picture for the behavior of the effective potential. So, in order to
draw consistent conclusions, we reconsider the equations of motion
of the fields $\phi_0$ and $\phi_1$,
\ba
\nabla^2\phi_0 &=& \frac{\lambda}{6}\,\phi_0^3+
\frac{\lambda}{2}\,(2N)\,\phi_0\,\phi_1^2
+\frac{\lambda}{2\sqrt{2}}\,(2N)^2\phi_1^3\\[3mm]
\nabla^2\phi_1 &=& \frac{\lambda}{2}\,\phi_0^2\phi_1 +
\frac{3\lambda}{2\sqrt{2}}\,(2N)\,\phi_0\,\phi_1^2 +
\frac{7\lambda}{12}\,(2N)^2\phi_1^3
\ea
By making use of the relation $\phi_0=q\,\phi_1$ and rearranging accordingly
the above equations, we obtain the following constraint for the
proportionality coefficient
\be
\frac{q^3}{3} +\frac{3\sqrt{2}}{2}\,N q^2 + \Bigl(\frac{7N^2}{3}-
N\Bigr) q -\frac{2N^2}{\sqrt{2}} =0\,\,.
\ee
This algebraic equation has the solutions
\be
q_1=-N\,\sqrt{2} \qquad , \qquad
q_{2,3}=\frac{-7N \pm \sqrt{49N^2+24N}}{2\sqrt{2}}\,\,.
\ee

When each one of the above values is substituted in the expression
(\ref{vfinal}), the potential exhibits a different behavior. 
Analytically:\\[4mm]
\underline{\bf (i) $q=q_1$}. In this case, we obtain:
\be
V_{eff}=\tilde{V}\,\Bigl(\frac{N}{N+1}\Bigr)
\label{veff1}
\ee
For $N=2$, this gives $V_{eff}=\frac 23\,\tilde{V}$. However, in
the limit $N \rightarrow \infty$, the effective potential asymptotically
tends to $\tilde{V}$. Note that the imposition of the periodic boundary
condition demands the existence of two boundaries so $N \geq 2$. 
When, at the next-to-lowest order, we studied the case $N=2$, we did not
obtain any solution with the coefficient that multiplies $\tilde{V}$
being smaller that unity. This means that the above solution owes its
existence to the imposition of the periodic condition on the Kaluza-Klein
fields (for $N = 2$) and it may not constitute a generic solution of the
original theory. However, for large $N$, we expect this behavior to
approximate the solution of the original Lagrangian.  In particular,
this is exactly the type of solution we were searching for.  Namely, at
large N, the potential of the late-time attractor fields does not depend
on $N$ relative to the original 4-dimensional potential.  Therefore, for
large $N$, chaotic inflation is realized in 4-dimensions with a quartic
5-dimensional coupling $\hat \lambda \sim 1$, and we have an explicit
example of assisted inflation.
\\[4mm]
\underline{\bf (ii) $q=q_2$}. Then, we have:
\be
V_{eff}=\tilde{V}\,\frac{2\,\Bigl[360+1904N+2401N^2-(156+343N)\,
\sqrt{49N^2+24N}\Bigr]}{(20+49N-7\sqrt{49N^2+24N})^2}
\ee
Then,
\ba
&~&{\rm For} \quad N=2 \quad : \quad V_{eff} \simeq 16.5\,\tilde{V}\\[3mm]
&~&{\rm For} \quad N>>2 \quad : \quad V_{eff} \simeq \tilde{V}\,
\Biggl\{7N+ \frac{18}{7} + {\cal O}\Bigl(\frac{1}{N}\Bigl)\Biggr\}
\label{veff2}
\ea
which clearly shows that the potential tends to increase with the number of
scalar fields that we include in the theory. The above solution corresponds
to the results (\ref{veffn=1}) and (\ref{veffn=2}) derived in the lowest
($N=1$) and next-to-lowest order ($N=2$), respectively. Both these solutions
showed a tendency to increase with the number of Kaluza-Klein fields, 
a behavior which obviously survived after the imposition of the periodic
condition. Of course this solution has exactly the $N$-dependence
that prohibits an assisted solution.\\[3mm]
\underline{\bf (iii) $q=q_3$}. In this case:
\be
V_{eff}=\tilde{V}\,\frac{2\,\Bigl[3640+1904N+2401N^2+(156+343N)\,
\sqrt{49N^2+24N}\Bigr]}{(20+49N+7\sqrt{49N^2+24N})^2}
\ee
Now,
\ba
&~&{\rm For} \quad N=2 \quad : \quad V_{eff} \simeq 1.04\,\tilde{V}\\[3mm]
&~&{\rm For} \quad N>>2 \quad : \quad V_{eff} \simeq \tilde{V}\,
\Biggl\{1+\frac{32}{343N} - \frac{368}{16807N^2} +
{\cal O}\Bigl(\frac{1}{N}\Bigl)^3\Biggr\}
\label{veff3}
\ea
In this case, the largest value that the potential takes on corresponds to
$N=2$ and, as we add more and more scalar fields, it asymptotically tends
to $\tilde{V}$ with the multiplication coefficient being always larger
than unity.  This solution is the analog of the solutions
(\ref{res1}), (\ref{res3}) and (\ref{res4}) derived in the next-to-lowest
order approximation.  By making
use of the periodic condition, we managed to include a large number of 
scalar fields in our model and found that these coefficients decrease
with the number of fields, as we expected. As in the case with $q = q_1$,
this class of solutions also allows for chaotic inflation with $\hat
\lambda \sim 1$ through assistance. 

Although the large number of fields has managed to remove the
fine-tuning problems, it is necessary to verify that the initial
conditions for inflation to occur are indeed fulfilled.  In four
dimensions, we normally assume $\tilde V(\tilde\phi) \sim M_P^4$, which
for $\lambda \ll 1$, corresponds to $\tilde\phi \gg M_P$. If these
conditions are translated into our five dimensional quantities, then we
would find, $\hat \phi \sim M_P^{1/2} M_5$ and $\hat V \sim M_P^2 M_5^3
\gg M_5^5$.  Without a better understanding of the dynamics of the
5-dimensional theory, we should instead insist that $\hat V(\hat \phi)
\sim M_5^5$. This condition, then, becomes
\be
\tilde V(\tilde \phi) = 2L \,\hat V(\hat \phi)
\sim M_P^2 M_5^2 < M_P^4
\label{infl1}
\ee
since $N=2LM_5=M_P^2/M_5^2 \gg 1$. The requirement
that the 4-dimensional coupling constant should be of ${\cal O}(10^{-12})$
imposes the following condition on the 5-dimensional coupling and
the four- and five-dimensional Planck mass
\be
\hat{\lambda}\,\left(\frac{M_5}{M_P}\right)^2 \sim 10^{-12}\,\,.
\ee
By appropriately choosing the values of the above quantities,
the required value of $\lambda$ is naturally obtained. However, the
initial condition for inflation $\tilde \phi \geq M_P$ puts a constraint
on the smallest  possible value of the ratio $M_5/M_P$~: when the above
condition is combined with the expression (\ref{infl1}) for the
4-dimensional potential, one finds
$M_5 \geq 10^{-6} M_P$. Then, even if $\hat \lambda$ is as large
as of ${\cal O}(1)$, we can still obtain $\lambda \sim 10^{-12}$. The
problems encountered when $M_5 \ll M_P$ have recently been discussed
\cite{ly,kl}. However, for $M_5 \sim 10^{-3}M_P$ as in many models of
string unification \cite{m}, we would have $N \sim 10^6$, and an initial
value of $\hat \lambda \sim 10^{-6}$ would be brought down to the
correct  four dimensional coupling. 

There is one more issue which must be addressed. In this section,
we have discussed the conditions leading to inflation, and the
attractor solution of the equations of motion. In doing so, we have 
neglected the KK mass terms, which is valid so long as $m^2 <
\lambda \phi_0^2$. At the onset of inflation, this condition is obeyed by
all of the KK fields only if the maximum mass we are considering (which
corresponds to the $N$th state and has mass $\sim M_5$) satisfies 
$M_5^2 < \lambda \phi_0^2 \sim \lambda^{1/2} M_P M_5$, or
$M_5 \la 10^{-6} M_P$. This means that only for the marginal value
of $M_5\simeq 10^{-6} M_P$ all of the KK fields can be considered
effectively massless while for $M_5\simeq 10^{-3} M_P$ we can ignore
the masses only for those fields with $m^2 < 10^{-3} M_5^2$. Moreover,
as the field $\tilde \phi$ moves toward the minimum of the potential,
$\phi_0$ becomes smaller as well and, gradually, more and more fields
cease to satisfy our assumption on the masses of the fields $\phi_i$. The
equations of motion of these massive fields are dominated by their mass
terms with the only late-time attractor being the trivial one. As a
result, these fields get decoupled from the rest of the system with a
time-scale inversely proportional to their mass: the more massive they
are, the faster they decouple. At the end of the day, when $\tilde \phi$
finally reaches the minimum of the potential, all of the massive KK
fields have decoupled and only the massless (by construction) KK
zero-mode, $\phi_0$, has remained playing  the role of the inflaton
$\tilde\phi$. However, this behavior does not affect the resolution of
the fine-tuning problem in the least. As the number of KK fields, that
can be considered massless, decreases, the solution $q=q_1$ gradually
disappears while the other two solutions, $q=q_2$ and $q=q_3$, tend to
become identical resulting in an effective potential which is again
independent of the number
$N$. As a result, the resolution of the fine-tuning problem holds at
all times: from the onset of inflation, when all or part of the KK fields
can be considered massless and contribute to the inflaton, until its final
stages, when all the KK fields have decoupled.
\paragraph{}
It appears that the compactification of a large extra dimension can lead
to assistance effects enhancing the probabilities for inflation not
only in the case of power-law potentials but in the case of exponential
potentials as well. As an illuminating example, we consider the
following 5-dimensional scalar field theory
\be
-{\cal L}_{5D}= \frac 12\,\partial_A \hat \phi\,\partial^A \hat \phi +
{\hat V_0}\,\exp\Biggl(-\sqrt{\frac{2}{\hat p}}\,
\frac{\hat\phi}{M_5^{3/2}}\Biggr)\,\,,
\ee
where $\hat V_0$ and $\hat p$ are constants. As in the case of the quartic
potential, the 5-dimensional field $\hat \phi$ (which is perhaps a
modulus field from the compactification of additional dimensions in the
theory) can be Fourier expanded along the compact coordinate
$z$. When the expansion (\ref{expansion}) is substituted in the above
Lagrangian, we obtain a scalar field theory described by
eq.~(\ref{trans}), interacting,
through an exponential potential. Even in terms of 5-dimensional
quantities, this theory is ladened with interactions since the potential
of every field multiplies the potential of every other field. When the
integration over the compact coordinate is conducted, we expect an
effective, heavily interacting, 4-dimensional scalar theory to arise.
However, the form of the potential makes the integration over
$z$ extremely difficult. Nevertheless, we can still make some qualitative
arguments on the assistance effect that follows from 
compactification. In terms of 4-dimensional quantities, the above
Lagrangian can be written as
\be
-{\cal L}_{eff}= \frac 12\,\partial_\mu \tilde \phi\,
\partial^\mu \tilde \phi +
\tilde V_0\,\exp\Biggl(-\sqrt{\frac{2}{\tilde p}}\,\tilde\phi\Biggr)\,\,.
\ee
where now
\be
\tilde{\phi}=\sqrt{2L}\,\hat\phi = \sqrt{N}\,\frac{\hat\phi}{M_5}
\quad , \quad
\tilde{V_0}=2L\,\hat V_0 = N\,\frac{\hat V_0}{M_5}\\[2mm]
\quad , \quad \tilde{p}=N\,\hat p\,\,.
\ee
As it is well known, a 4-dimensional theory of the form given above leads
to a power-law
expansion of the Universe: $R(t) \sim t^{\tilde p}$. After 
compactification, the parameter $\hat p$ has been multiplied by the number
of massive KK fields that are present in the theory and, as a result, for
sufficiently large $N$, the 4-dimensional theory will produce inflation
even if the 5-dimensional theory with the parameter $\hat p$ was not
able to. Finally, it is worth noting that the above dependence
of the field $\tilde\phi$ and the parameters $\tilde V_0$ and $\tilde p$
on the number of multiple fields $N$ was also derived in~\cite{liddle}
although the origin of the fields was not specified. Here, we argue that
the compactification of a 5-dimensional theory with an exponential potential
could provide us with both the necessary multiplicity of scalar fields
and the desired dependence of the parameters of the theory on the number
of fields. 

\paragraph{}
We summarize the results of this section: The Kaluza-Klein
compactification of the fifth dimension of a 5-dimensional theory of a
single, self-interacting scalar field leads  to the appearance of a large
number of Kaluza-Klein scalar fields in the 4-dimensional effective
theory. A feature of this effective theory is the presence of a
complicated web of interaction terms between the scalar fields of the
theory. Once the late-time attractor of the system is determined, this
field theory of multiple scalar fields can be mapped to a theory of a
single, self-interacting scalar field $\tilde{\phi}$. The presence of the
interaction terms drives the effective potential towards two different
directions: in one case, it increases with the number of extra scalar
fields that are present in the theory while, in the second case, is
starts with a value slightly smaller or larger than the value of the
one-field-self-interaction potential
$\tilde{V}$  but asymptotically tends back to $\tilde{V}$. At the end of
the analysis, the renormalized scalar field $\tilde{\phi}$ turns out to be
much more ($q = q_2$) or equally strongly coupled ($q = q_{1,3}$) compared
to the initial 4-dimensional Kaluza-Klein fields. As a result, the
renormalized coupling $\tilde{\lambda}$, defined as $\lambda$ multiplied
by the expressions in brackets in eqs.~(\ref{veff1}), (\ref{veff2}) and
(\ref{veff3}), takes on a value which is much larger ($q = q_2$) or almost
the same ($q = q_{1,3}$) compared to the value of the initial 4-dimensional
coupling $\lambda$. However, in the latter cases, $\tilde{\lambda}$ is
suppressed by the number of scalar fields relative to the original coupling
$\hat{\lambda}$ of the 5-dimensional scalar field.  This a concrete example
of assisted inflation.

\section{Conclusions}

In this paper, we have dealt with the problem of fine-tuned coupling
constants in the framework of field theories that involve self-coupled or
interacting scalar fields. This problem inevitably arises when we consider
the possibility of the creation of an inflationary epoch in our universe
and demand an agreement between the theoretical predictions and the 
experimental (COBE) data on density fluctuations.

We have demonstrated by considering some general field theories of
multiple scalar fields in 4 dimensions  that the idea of assisted
inflation based on exponential potentials~\cite{liddle} can be easily
extended in the case of power-law potentials. In this case,
the pre\-sence of multiple scalar fields leads to a renormalized theory
of a single scalar field which is considerably less strongly coupled
than the original fields of the theory. The renormalized coupling
constants scale with the number of fields $N$ which permits the
creation of an inflationary period without severe fine-tuning.
However, the effectiveness of assistance depends strongly
on the interactions between the scalar fields of the theory.
If the multiple scalar fields are assumed to be only self-coupled,
both power-law inflation based on exponential potentials as well as 
chaotic inflation  works well with only mild or no fine-tuning at all
depending on the number of fields $N$ that we include in the theory. If,
on the other hand, we allow cross-coupling terms between different scalar
fields, the assistance method breaks down
leading to a much more strongly coupled theory. 

As a concrete example of a field theory with multiple scalar fields,
we considered a single, self-interacting 
scalar field living in 5 dimensions with a quartic potential. 
(Other recent constructions for inflationary models involving a large
extra dimension can be found in refs. \cite{nem,ly,kl,ikko,ekoy,ov}.)
Assuming that the fifth dimension is compactified along a circle and
applying a Kaluza-Klein reduction of the 5-dimensional field, we obtained
a 4-dimensional, effective theory with the necessary multiplicity of
scalar fields fulfilled by the presence of the Kaluza-Klein modes. The
resulting potential contains a complex network of cross-coupling terms. As
suggested by our previous results, these interaction terms are expended
to hinder inflation. In terms of 4-dimensional quantities, this is indeed
the case. Once the theory of multiple Kaluza-Klein fields is mapped to a
theory of a single, renormalized scalar field, we found three different
solutions for the corresponding effective potential: the first one
follows a behavior similar to the one derived in the purely
4-dimensional case and drives the potential, and thus the renormalized
coupling constant, to large values increasing with the number of scalar
fields; the other two solutions start with a value for the effective
potential which is slightly smaller or larger than the value of the
one-field-potential
$\tilde{V}$ but asymptotically tends to $\tilde{V}$ as we increase the
number of fields. In both cases, the desired behavior of the effective
potential is not achieved and the renormalized scalar field is either
more or equally strongly coupled than the original 4-dimensional
Kaluza-Klein fields. Consequently, the necessary fine-tuning of the
renormalized quartic coupling constant 
$\tilde{\lambda}$ becomes more severe or at best remains the same compared
to that of $\lambda$, a result which is attributed to the presence of
interaction terms between the Kaluza-Klein fields of the theory. 
However, the theory of the renormalized, scalar field does indeed get
assisted although via a different path. The 4-dimensional coupling
constant $\lambda$ of the Kaluza-Klein fields is determined by the
5-dimensional one, $\hat{\lambda}$, divided by the number of the 
Kaluza-Klein modes. As a result, $\lambda$ and thus $\tilde{\lambda}$, 
in the case of the latter two solutions, is suppressed by the number
of scalar fields that are present in the theory relative to the 
original coupling constant $\hat{\lambda}$ of the 5-dimensional theory. 
By choosing appropriate values of the five-dimensional
Planck mass $M_5$ and the five-dimensional coupling constant $\hat\lambda$,
we are able to naturally obtain a four-dimensional, self-interacted
scalar theory with $\lambda \sim {\cal O}(10^{-12})$ (in
agreement with COBE data) without the need of any
fine-tuning. Moreover, our results do not depend on the number of
massive KK fields that contribute to the inflaton field and, as a result,
the resolution of the fine-tuning problem holds from the onset of
inflation until its final stages.
\vspace*{4mm}

{\large\bf Acknowledgments}
We would like to thank Nemanja Kaloper for many useful and enlightening
discussions. This work was supported in part by 
DOE grant DE-FG02-94ER40823 at Minnesota.

\end{document}